\documentclass{article}
\usepackage{fullpage}
\usepackage{amsfonts,amssymb}
\usepackage{amsthm}
\usepackage{tikz}
\usetikzlibrary{calc}
\usetikzlibrary{arrows}
\usepackage{tikz-3dplot}
\usepackage{color} 
\usepackage[fleqn]{amsmath}
\usepackage{amsthm}
\usepackage{amssymb}
\usepackage{amsfonts}
\usepackage{bbm}
\usepackage{epsfig}
\usepackage{times}
\usepackage{hyperref}
\usepackage{bm}
\usepackage{float} 
\usepackage{appendix} 
\usepackage{lscape}
\usepackage{cite}
\usepackage{mathrsfs}
\usepackage{appendix}
\usepackage{setspace}
\usepackage{mathtools}
\usepackage{mdframed}
\usepackage{authblk}
\theoremstyle{definition}

\usepackage{tikz}

\begin{document}

\title{Two Roads to Retrocausality }

	\author[1]{Emily Adlam} 
	\date{\today} 
	\affil[1]{The University of Western Ontario} 
\maketitle

In recent years the quantum foundations community has seen increasing interest in the possibility of using retrocausality as a route to rejecting the conclusions of Bell's theorem and restoring locality to quantum physics\cite{RevModPhys.92.021002,article,Miller1996RealismAT}. On the other hand, it has also been argued that \emph{accepting} and embracing nonlocality also leads to a form of retrocausality\cite{Adlamspooky}. It is interesting that two diametrically opposite starting points both seem to lead to the same conclusion, suggesting that the relationship between retrocausality and locality is a complex one. In this article we seek to elucidate that relationship and draw some conclusions about the most appropriate route to retrocausality. 

We begin by providing a brief schema of the various ways in which violations of Bell's inequalities might lead us to consider some form of retrocausality. 
We then consider some possible motivations for using retrocausality to rescue locality, arguing that none of these motivations is adequate and that therefore there is no clear reason why we should prefer local retrocausal models to nonlocal retrocausal models. Next, we examine several different conceptions of retrocausality, concluding that `all-at-once' retrocausality is more coherent than the alternative dynamical picture. We then argue that since the `all-at-once' approach requires probabilities to be assigned to entire histories or mosaics, locality is somewhat redundant within this picture. Thus we conclude that using retrocausality as a way to rescue locality may not be the right route to retrocausality.

Finally, we demonstrate that accepting the existence of  nonlocality and insisting on the nonexistence of preferred reference frames leads naturally to the acceptance of a form of retrocausality, albeit one which is not mediated by physical systems travelling backwards in time. We argue that this is the more natural way to motivate retrocausal models of quantum mechanics.

At this juncture we specify that in this article we are using the term `retrocausality' in a fairly general way which is intended to cover any metaphysical picture in which future events are regarded as having some sort of influence on past events.  One might legitimately have questions about whether all such influences are `causal' in nature, and of course proponents of different analyses of causation are likely to take different views on this question. It's possible that there is simply no definite answer as to whether or not these sorts of  influences are `causal' -  this is a question about how best to extend our existing concept of causation to a new domain, and often when concepts are extended to a new domain there is often some freedom of choice about how exactly we wish to apply them within that new domain. But we will nonetheless make use of the term `retrocausality' throughout this article to refer to such backwards influences, since that is the language most commonly employed in the relevant literature. 

Note that in addition to arguments concerning locality, there has been a substantial literature on the connection between retrocausality and time symmetry\cite{2010arXiv1002.0906P,PuseyLeifer}. We do consider that  arguments using time symmetry to motivate retrocausality have merit, but for reasons of space we will not discuss such arguments in this article.

\section{Bell correlations and Retrocausality}

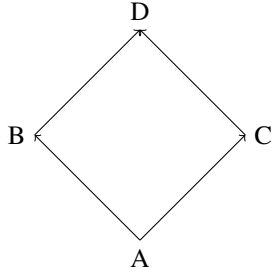
\begin{figure}
	\begin{center}
		
		\begin{tikzpicture}[scale=0.7]
		

		\draw[->] (0,0) -- (2,2) ;
		\draw[->] (0,0) -- (-2,2) ;
		\draw[->] (-2,2) -- (0,4) ;
		\draw[->] (2,2) -- (0,4) ;

		\node [below  ] at (0,0) {A};
		\node [right  ] at (2,2) 	{C};
		\node [left  ] at (-2,2) {B};
		\node [above] at ( 0,4) {D};
		
	\end{tikzpicture}

\caption{Schematic experimental geometry for the simple Bell experiment discussed throughout this article. $A$ is the spacetime point where the Bell state is prepared, $B$ and $C$ are the spacetime locations in the future lightcone of $A$ where the two measurements are performed on the two particles from the Bell pair, and $D$ is some spacetime location in the future lightcone of both $B$ and $C$ - for example, perhaps the location where the results of the measurements at $B$ and $C$ are compared. }

\label{prep}
\end{center}
\end{figure}

Bell's theorem tells us that, subject to certain assumptions, any  realistic, intersubjective account of quantum phenomena must explain the relationships between the correlated outcomes of measurements on a Bell pair in terms of some mechanism other than a common cause in the past lightcones of the measurements. So, assuming we believe the assumptions that go into Bell's theorem and we are not willing to give up on the attempt to provide a realistic, intersubjective account of quantum phenomena, we seem to have three options, which can be seen clearly on the diagram in figure \ref{prep}, where we suppose that the experimenters at $B, C$ make free choices of their measurements settings at the time of their measurements,

\begin{enumerate}  \item The beables mediating the correlations are not located along any continuous spacetime pathway between $B$ and $C$, or there are no beables at all mediating the correlations
	
	\item  The beables mediating the correlations are located along a continuous, partly spacelike (faster-than-light) pathway between $B$ and $C$
	
	\item  The beables mediating the correlations are located along a continuous, lightlike or timelike and partly retrocausal pathway between $B$ and $C$ which goes via the future (e.g. point $D$) or the past (e.g. point $A$). 
	
	\end{enumerate} 

Option 2) is usually cashed out in terms of something like a `collapse of the wavefunction' (the collapse will be instantaneous in one frame of reference and thus will appear to occur at finite but faster-than-light speeds in some other frames of reference). Because we  have to postulate a preferred reference frame on which the wavefunction collapses, this route is somewhat in tension with relativity. Moreover it has been pointed out that the hypothesis that nonlocal influences are mediated by some finite superluminal process makes different predictions from standard quantum mechanics in certain special cases\cite{Bancal_2012}. Actually performing the experiments to differentiate between these hypotheses is not straightforward, but so far all the evidence points to the predictions of quantum mechanics being correct. For these reasons, we will not discuss option 2) further in this article.

Option 1) involves accepting the existence of   unmediated nonlocal influences. In section \ref{myway} we will see that combining this view with relativistic constraints (i.e. demanding the nonexistence of a preferred reference frame) leads naturally to a kind of retrocausality. But for now we will mainly focus on option 3), which of course explicitly involves accepting the existence of retrocausality. 
 
As an alternative to these three options,  one could also consider  denying one of the assumptions that goes into Bell's theorem, thus potentially restoring the possibility of explaining the Bell correlations in terms of a common cause in the past lightcone.  There are a variety of such assumptions, but here we will confine our attention to those which bear some relation to retrocausality (so we will not, for example, discuss the many-worlds approach, which denies the assumption that measurements always have a single outcome). One assumption which has received significant attention in recent years is   `statistical independence,' i.e. the assumption that the ontic state prepared at $A$ is independent of the choices of measurement settings made at points $B$ and $C$. Approaches which deny statistical independence are often known as `superdeterministic,'\cite{hossenfelder2020superdeterminism,10.3389/fphy.2020.00139} although in fact models of this sort do not necessarily have to be deterministic\cite{adlam_2021}. There are three possible ways to explain violations of statistical independence: 

\begin{enumerate} 
	
	\item The experimenters' choices at $B$, $C$ are caused or non-causally influenced by the value of the ontic state at $A$. 
	
	\item The value of the ontic state at $A$ is caused or non-causally influenced by the experimenters' choices at $B$ and $C$. 
	
	\item Both the experimenters' choices at $B$, $C$ and the local hidden hidden variables have some joint common cause or non-causal explanation. 
	
\end{enumerate} 

Of these options, 2) transparently requires that events at $A$ depend on events at $B$ and $C$, so it is a retrocausal  approach in the general sense in which we are using that term. 
Moreover, we can easily arrange for option 1) to exhibit similar features: we may require that  the experimenters make their choices some time before the experiment begins, so those choices are made in the past lightcone of $A$, and thus, if we assume that the underlying mechanism works the same in both cases, we would then have a retrocausal influence from $A$ to the location of the choice. (In sections \ref{symmetry}, \ref{myway} we will examine in greater detail the assumption that the underlying mechanism works the same in both cases).  Thus neither of these approaches straightforwardly delivers on the original vision of superdeterminism as a way of explaining the Bell correlations purely in terms of common causes in the past lightcone of the relevant measurements. However,  since we are free to suppose that the influence of the measurement choice on the ontic state is mediated by a retrocausal process involving a continuous spacetime pathway, there is a sense in which locality is maintained.   Option 3) on the other hand allows us to avoid retrocausality altogether and thus models of the third kind  may offer a way of maintaining locality without retrocausality. The \emph{Invariant Set Theory} approach is making some interesting progress in this direction\cite{palmer2016invariant}, but since this is not a retrocausal approach we will not address it further in this paper.

Thus we see that a number of different proposals for realist, intersubjective accounts  of the Bell correlations lead to some form of retrocausality, broadly construed. However,  there are two importantly different routes here. Either we use retrocausality to rescue locality, and thus account for the Bell correlations without the need for a preferred reference frame, or we explicitly accept nonlocality and then it transpires that demanding the nonexistence of a  preferred reference frame leads us toward a form of retrocausality in any case. It's clear, therefore, that the combination of the Bell correlations with the absence of a preferred reference frame is somehow linked with retrocausality, but what role does locality play in all this? For those of us who are willing to accept the existence of retrocausality, should we be using it to rescue locality or should we just accept nonlocality and retrocausality as part of a single package?

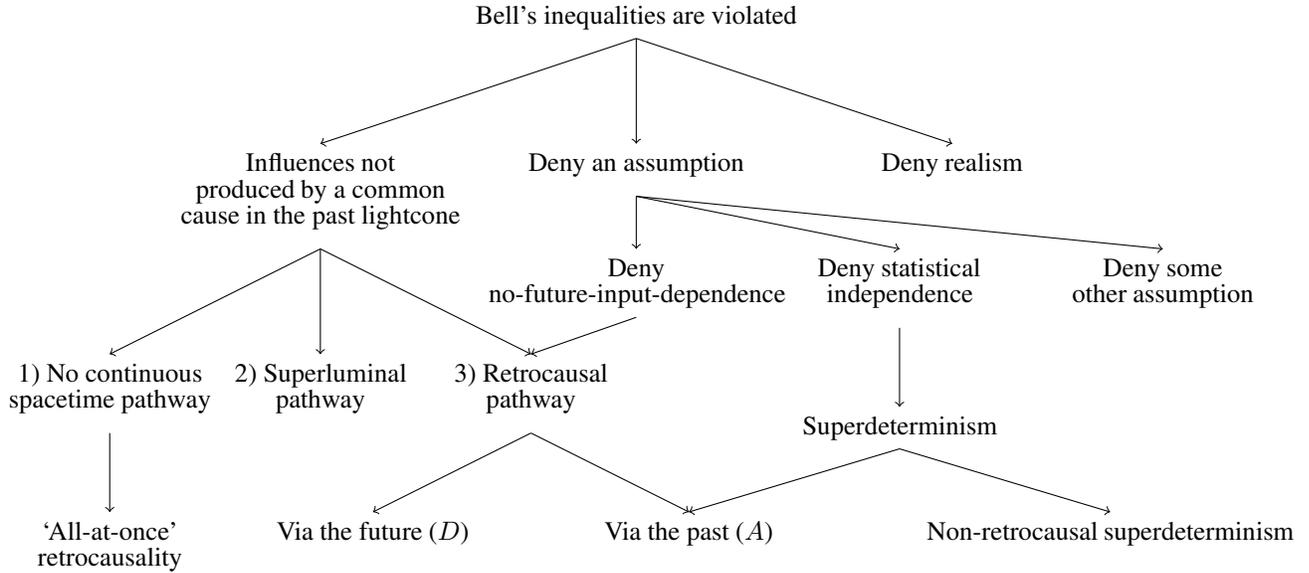
\begin{figure}
	\begin{center}
		
		\begin{tikzpicture}[scale=0.7]
		
		
				\node [above ] at (0,0) {Bell's inequalities are violated};
		\draw[->] (0,0) -- (-6,-2) ;
		\draw[->] (0,0) -- (0,-2) ;
		\draw[->] (0,0) -- (6,-2) ;

		\node [below  ] at (-6,-2) {Influences not};
		\node [below  ] at (-6,-2.5) 	{produced by a common};
		\node [below  ] at (-6,-3) {cause in the past lightcone};
		\node [below] at ( 0,-2) {Deny an assumption};
		
		\node [below ] at (6,-2) {Deny realism};

		\draw[->] (0,-3) -- (5,-4) ;
		\node [below ] at (5,-4) {Deny statistical} ;
		\node [below ] at (5,-4.5) {independence};

		\draw[->] (5,-5.5) -- (5,-7) ;
		\node [below ] at (5,-7) {Superdeterminism} ; 
		
		\draw[->]  (5,-7.8) -- (1,-9) ;
		
		\draw[->]  (5,-7.8) -- (9,-9) ;
		
		\node [below ] at (9,-9) {Non-retrocausal superdeterminism} ;
		
		\draw[->] (0,-3) -- (0,-4) ;
		\node [below ] at (0,-4) {Deny } ;
		\node [below ] at (0,-4.5) {no-future-input-dependence};
		
		\draw[->] (0,-3) -- (10,-4) ;
		\node [below ] at (10,-4) {Deny some} ;
		\node [below ] at (10,-4.5) {other assumption};
		
		\draw[->] (0,-5.3) -- (-2,-6) ;

		\draw[->] (-6,-4) -- (-10,-6) ;
		\draw[->] (-6,-4) -- (-6,-6) ;
		\draw[->] (-6,-4) -- (-2,-6) ;

		\node [below  ] at (-10,-6)  {1) No continuous};
		\node [below  ] at (-10,-6.5)  {spacetime pathway};
		
		\draw[->] (-10,-7.5) -- (-10,-9) ;
		\node [below, ] at (-10,-9) {`All-at-once'};
		\node [below ] at (-10,-9.5) {retrocausality};

		\node [below] at(-6,-6)  {2) Superluminal};
		\node [below] at(-6,-6.5)  { pathway};
		\node [below ] at (-2,-6) {3) Retrocausal};
		\node [below ] at (-2,-6.5) { pathway};

		\draw[->] (-2,-7.5)  -- (-5,-9) ;
		\draw[->]  (-2,-7.5) -- (1,-9) ;
		
		\node [below  ] at  (-5,-9) {Via the future ($D$)};
		\node [below] at (1,-9) {Via the past ($A$)};

		\end{tikzpicture}
		
		\caption{A schematic diagram setting out various possible ways of responding to the violations of Bell's inequalities and the relationships between them}
		
		\label{prep}
	\end{center}
\end{figure}

\section{Motivations for Rescuing Locality  \label{reasons}}

The idea that retrocausality can be used to rescue locality has a long history. It seems to have been proposed first by Costa de Beauregard\cite{deBeauregard1953-DEBMQ,deBeauregard1976-DEBTSA,CostaDeBeauregard:1977ik}, and has since been taken up by a number of physicists, with contemporary proponents including  Price\cite{PRICE_1994}, Wharton\cite{Wharton_2018}, Sutherland\cite{Sutherland:1983wq,article}, Evans\cite{Evans_2013}, Miller\cite{Miller1996RealismAT} Argaman\cite{RevModPhys.92.021002} and others. A variety of explicit retrocausal models have been developed, including Schulman's anomaly models\cite{schulman1997time,e14040665} and Wharton's Lagrangian models\cite{WhartonKleinGordon,2016hwar}; perhaps the most fully developed retrocausal approach at present is the transactional interpretation proposed by Cramer\cite{PhysRevD.22.362,article} and Kastner\cite{doi:10.1063/1.4982766}, which describes experiments in terms of `offer waves' travelling forwards in time produced by emitters (e.g. preparation devices) and `confirmation waves' travelling backwards in time produced by absorbers (e.g. measurement devices), which interact to form a transaction and thus actualise some  outcome of the relevant measurement. 

The motivation for the retrocausal response to Bell's theorem is quite simple. One of the assumptions going into the derivation of Bell's theorem is `no future input-dependence.' ie `\emph{no model parameter associated with time $t$ can be dependent upon model inputs associated with times greater than $t$}.' \cite{RevModPhys.92.021002} So if we allow this assumption to be violated, we can come up with a realistic, intersubjective model of the Bell experiment in figure \ref{prep} which exhibits a form of locality - we simply postulate an influence mediated by some beable which is located along a continuous lightlike or timelike path which goes forward to point $D$ and then back into the past, or alternatively goes back to point $A$ and then forward into the future.  As discussed by Wharton and Argaman in ref \cite{RevModPhys.92.021002}, the resulting models will still violate Bell's assumption of local causality, because local causality as defined by Bell is the conjunction of no future-input dependence with a screening assumption, but nonetheless `\emph{an essential element of `locality' can be retained}' as these models satisfy a condition that Wharton and Argaman refer to as \emph{continuous action} (CA): that is to say, these models utilize spacetime-based parameters associated with intermediate regions between preparations and measurements and therefore they  are `locally mediated,' i.e. correlations cannot be introduced or altered except via intermediate spacetime-based mediators. The possibility of rescuing locality in this specific sense is often cited as a key motivation for moving to retrocausal models - see for example refs \cite{RevModPhys.92.021002,Priceretro,Evans_2013}.

Approaches making use of retrocausality to restore locality in the CA sense are evidently motivated by a conviction that local theories are preferable to nonlocal ones. And this is indeed a widely held view - even now many physicists are very reluctant to accept the existence of genuine unmediated nonlocal influences. But in this section, we will argue that retrocausal theories satisfying CA do not accrue any of the benefits one might hope to gain from local theories. Specifically, our conclusion will be that anything which can be achieved by a retrocausal model satisfying CA can also be achieved within an equivalent retrocausal model which does not satisfy CA, and therefore considerations of simplicity and ontological economy give us good reason to do away with these intermediate beables.

 For simplicity, we will proceed by making a comparison between two specific models. $M_1$ is some model which obeys CA, i.e. in $M_1$ all correlations are mediated by beables which are located along continuous spacetime paths connecting all the relevant events. For example $M_1$ could be a version  of the transactional interpretation\cite{Cramer}, or one of Wharton's path integral models\cite{Wharton_2018}. Meanwhile, $M_2$ is identical to $M_1$ except that there are no beables in the regions between the events: in this model the relations between the events are understood as purely a form of objective modal structure. That is to say, in $M_2$ the modal relations between the events (perhaps causal relations, or perhaps some other sort of modal relation such as ontological priority or metaphysical necessitation) exist and are instantiated by the events in question without any need for mediation via some physical bearer of modal powers; $M_2$ therefore belongs to the class of modal interpretations of quantum mechanics\cite{Vermaas1999-VERAPU}. We reinforce that $M_1$ and $M_2$ are not mathematically different, but they differ at the level of ontology and interpretation. As we will discuss later, removing the CA requirement significantly expands the space of possible models and therefore we would expect that in the long term moving away from CA would lead to new models which are mathematically different from existing CA models, but for the purpose of making a clear comparison it is most straightforward to consider two models which differ only in what they say about the existence of intermediate beables.

 \subsection{Consistency with relativity \label{reason1}} 
 
 The most common reason given  for attempting to rescue CA is that this allows us to make quantum mechanics consistent with relativity\cite{RevModPhys.92.021002,Priceretro,Cramer}. This point is motivated by the fact that most extant explicitly non-local interpretations of quantum mechanics, such as the de Broglie-Bohm interpretation and the GRW collapse model,  model non-locality via a collapse or update of the wavefunction which occurs instantaneously everywhere, and thus we usually have to define a preferred reference frame on which this collapse or update can occur.  The existence of this preferred reference frame does not lead to outright contradictions with relativity in the de Broglie-Bohm or GRW models, because it turns out that the preferred reference frame can't be directly observed and thus it will not lead to any experimental predictions which are in conflict with the predictions of relativity\cite{Passon2006-PASWYA}. However, reintroducing a preferred reference frame looks like a major step backwards\cite{cushing1994quantum}, since a large part of the scientific revolution that resulted from special relativity was due to Einstein's rejection of the classical notion of `absolute space' and the associated notion of preferred reference frames. Moreover one important reason for being interested in the interpretation of quantum mechanics is the hope that getting clear about interpretative questions might help with  ongoing efforts to unify quantum mechanics with general relativity\cite{hardy2010physics}, and  an approach which so explicitly runs counter to the spirit of relativity doesn't seem likely to be productive on that front. 
 
 Thus the goal of accounting for the Bell correlations without introducing  a preferred reference frame is certainly a very reasonable one. And indeed, models like $M_1$ which maintain continuous action by placing mediating beables along a zig-zag path through the future or the past do achieve that goal, since a continuous timelike or lightlike path through spacetime is a relativistically covariant object and therefore no preferred reference frame is required for these models. However, the relativistic covariance of these models has nothing to do with the fact that they postulate \emph{mediation}:  consistency with relativity follows immediately from the fact that the models don't make use of an instantaneous collapse or update, so the mediating continuous paths really have nothing to do with the relativistic covariance. Indeed,  the alternative model $M_2$ which gets rid of the continuous paths and allows the preparation and measurement events to depend directly on one another is also relativistically covariant, because it likewise refrains from postulating an instantaneous collapse or update.  So the mandate to avoid preferred reference frames does not in and of itself give us any particular reason to prefer the continuous paths model $M_1$ to the direct dependence model $M_2$.   Moreover, Ockham's razor would seem to favour $M_2$ over $M_1$ - given that the mediating beables aren't really needed to maintain consistency with relativity, we might as well get rid of them.

 \subsection{Explaining the apparent locality of the classical world \label{reason2}} 
 
Another possible motivation is simply the desire to preserve our classical intuitions about reality: we experience the world as local, so we would like to have an interpretation of quantum mechanics which is also local. But this seems like a fairly weak argument: quantum mechanics has already forced us to discard many of our classical intuitions, so why should locality be any different? 

A better version of this argument starts with the observation that much of reality \emph{appears} to be local - for example, it isn't possible for us to influence distant objects without employing a physical process which propagates via a continuous path in spacetime.  This is a striking feature of the physical world which seems to demand an explanation - and one might reasonably think that the best way to explain this feature is to postulate that the world really is local at the fundamental level.  Moreover, throughout the history of science locality principles  have in general been highly successful   - for example, quantum field theory is based on a principle known as `locality,' which in different formulations of the theory refers either to the fact that spacelike separated operators commute, or that we employ only Lagrangians in which all interaction terms apply to fields at the same spacetime point\cite{Tongqft}. The fact that QFT obeys `locality' in these senses doesn't entail that it is local in either the Bell sense or the CA sense, and indeed,  it has been shown that violations of Bell inequalities are generic in QFT\cite{brunetti2012locality}. But nonetheless, the fact that these specific locality principles have worked so well in the formulation of QFT is something which any interpretation of quantum mechanics should seek to explain, and surely the most obvious explanation is to say that the world really \emph{is} local at the most fundamental level?

 But this argument too has problems. For the retrocausality employed in $M_1$ and other similar CA-satisfying models is explicitly intended to allow us to create an appearance of nonlocality under certain circumstances, in order to explain violations of Bell inequalities. And therefore prima facie there seems no reason why this sort of retrocausality could not be used to create the appearance of \emph{much stronger} nonlocality whilst still satisfying $CA$.  For example, in a model allowing retrocausality with no restrictions on what sorts of backwards influences we can create, it would be possible to produce the appearance of superluminal signalling in a Bell experiment by sending a signal to the future and then back into the past (e.g. via point $A$ in figure \ref{prep}). Indeed,  in a model allowing unrestricted retrocausality it would be possible to make phenomena appear almost as nonlocal as one could possibly imagine by simply sending signals very far into the future and then back into the past. Yet all of these models would still obey $CA$, since the signals are transmitted via continuous spacetime paths. So in fact, simply postulating CA together with retrocausality as in the model $M_1$ does nothing to explain the appearance of locality in the macroscopic world. 
 
In order to provide such an explanation, $M_1$ needs some principled way to limit the sorts of retrocausal influences that actually obtain in reality, which thereby places limits on the extent to which retrocausal influences can be used to create an appearance of nonlocality. For example,   ref \cite{2015arXiv151003706A} shows that Schulman's retrocausal model\cite{schulman1997time} prevents signalling backwards in time by imposing a  symmetry on the model which has the effect that the different possible retrocausal influences are equivalent from the point of local observers. It is these  sorts of restrictions on the set of possible retrocausal influences in $M_1$, rather than the fact that it obeys $CA$, which give rise to the appearance of locality at the macroscopic level; and since $M_2$ is by definition subject to all of the same constraints on the set of possible retrocausal influences, it will give rise to the appearance of locality in just the same way even though it explicitly does not obey CA. So CA is not actually doing any work in model $M_1$ to explain the apparent locality of the macroscopic world or the success of locality principles: all the work is done by the set of restrictions on allowed retrocausal influences, and thus $M_2$ can achieve equal explanatory power without postulating mediation or continuous action.

\subsection{Ontology \label{ontology} } 
 
 Another  kind of motivation for CA is based on considerations of ontology. If for example we want to adopt a particle-based ontology, then naturally we would like to account for Bell correlations with a model obeying CA, as  a defining feature of particles  is that they travel along continuous spacetime pathways. The same goes for an ontology based on waves or fields, since classical wave theories and all our successful field theories describe behaviour which is local in spacetime.  So plausibly one might be attracted to models satisfying CA in order to facilitate the adoption of a familiar ontology. 
 
 But this seems to be getting things the wrong way round: we should not draw conclusions about reality based on our presuppositions about ontology, rather we should make inferences about ontology based on what conclusions our theories lead us to draw about reality.  Of course, it  may be argued that we \emph{have} in fact inferred the existence of particles and/or waves and/or fields from our theories, because we arrive at such things by reifying certain elements of the mathematical structure of our theories - for example, by assuming the quantum state is an element of reality, we arrive at an ontology based on a wave or field evolving unitarily and locally.  But this claim is  questionable.  First, we know that the locally evolving wavefunction can't be the whole story, because we still need to make sense of the collapse of the wavefunction - and if we have ruled out the idea of a literal collapse or state update because we don't wish to introduce a preferred reference frame, it seems we must either become Everettians\cite{Wallace} or take the wavefunction itself less literally. Second, there are often several distinct mathematical descriptions for quantum processes, so how do we know which one we ought to reify in this way? Why for example should we base our ontology on temporally evolving states (the central object of the Schr\"{o}dinger picture) rather than temporally evolving operators (the central object of the equivalent Heisenberg picture)?\cite{sakurai2020modern}

Furthermore, consider for a moment the plight of some hypothetical observers who live in a world which does in fact generically violate CA, i.e. a world in which events stand in modal relations to one another without any mediating beables in the spacetime regions between the events.  When the observers try to express the nature of these modal relations mathematically for the purpose of predicting future events, their mathematical description will naturally have to include mathematical structures of some sort - differential equations, Lagrangians, Hamiltonians, operators or whatever else their mathematicians might dream up. And since these structures relate events which are located in spacetime, it's likely that the observers would be able to find a way of interpretating parts of these structures as some sort of beables occupying the spacetime regions between the events in questions. But ex hypothesi in this world these relations are not mediated by anything physical; so the mere fact that our theories contain mathematical structures which could potentially be interpreted as mediating spacetime processes does not entail that there really \emph{are} any mediating spacetime processes, and therefore we still need some specific reason to prefer $M_1$'s  interpretation in terms of mediating processes over $M_2$'s interpretation in terms of pure modal structure.   Indeed, we already know that there are significant obstacles to interpreting the mathematical structures of quantum mechanics as physical beables which satisfy CA - for example, the wavefunction lives on configuration space rather than ordinary spacetime, so it can't straightforwardly be interpreted as an ordinary CA-satisfying physical field as in model $M_1$, whereas if the wavefunction is regarded as simply a mathematical expression of modal structure as in model $M_2$, then it is not at all surprising that it does not live on physical space.  

 Another type of ontological objection involves the concern that it's unclear what ontology we are left with if we remove intermediate beables from our models.  For example, the transactional interpretation describes a process of emission and absorption of offer and confirmation waves which takes place between `emitters' and `absorbers,' which are typically assumed to be macroscopic objects\cite{Cramer}: if we get rid of the intermediate quantum waves and transactions, we are left with just the `emitters' and `absorbers,' but what are emitters and absorbers made of if not quantum systems?   Similarly, in Wharton's field-based retrocausal models, \emph{`External measurements (both before and after the subsystem in question) will be treated as physical constraints, imposed on the subsystem in exactly the same way that boundaries are imposed when using Hamilton’s principle,'}\cite{sym2010272} so the model says nothing about the nature of the preparation and measurement devices which establish the relevant constraints, and therefore it's not  clear what ontology remains if we get rid of the intermediate fields and postulate direct relations between these preparation and measurement devices. 
  
 This objection has some bite - the lack of clarity around the ontology of emitters, absorbers, preparation and measuring devices and so on is certainly a problem for these retrocausal interpretations. But exactly the same problem arises even if we retain mediating beables in our ontology:  these models rely on a well-defined distinction between the microscopic (the mediating beables) and the macroscopic (the emmitters/absorbers/etc) which we are apparently required to accept as a primitive, and yet the fact that conventional quantum theory treats the distinction between the microscopic and the macroscopic as primitive is one of the main reasons why it needs an interpretation in the first place! Thus whether or not we take the mediating beables seriously as part of our ontology, it remains the case that without answers to these questions the retrocausal approaches don't represent a complete interpretation of quantum mechanics. So there is certainly work to be done to establish a well-defined ontology for these approaches which explains what measuring devices and macroscopic objects actually are, but this work must be done regardless of whether or not we include mediating beables in our ontology:  removing the propagating waves or particles from our ontology makes the problem more noticeable, but the difficulty was really there all along and therefore these considerations are not in and of themselves a reason to insist on continuous action.

 \subsection{Symmetry Arguments \label{symmetry}}
 
 A novel argument for the CA approach is presented in ref \cite{Evans_2013}. Here, the authors discuss an experiment (SEPRB) in which the experimenter chooses from one of two ways of inputting a photon to a polariser, and the photon is then passed through a second polariser and measured. Obviously here the second measurement must be at a timelike separation from the location of the experimenter's initial choice. The authors note that the correlations between the experimenter's choice of polarizer setting and the result of the final measurement are identical to the correlations obtained between the choice of measurement on one side the the measurement outcome on the other in the case of entangled Bell particles, as in figure \ref{prep} (EPRB). They argue that since the correlations are identical in both cases, we should assume that the underlying mechanism producing the correlations is of the same kind in both cases. But of course, in the SEPRB case we typically suppose that the correlations are the result of information carried in the state of the photon as it travels along a continuous spacetime path from the first measurement to the second measurement, i.e. our usual model of these correlations satisfies CA. So the authors conclude that in the EPRB case we should likewise suppose that the correlations are the result of information carried in the state of some beable as it travels along a continuous spacetime path from one measurement to the other measurement, and thus this case also satisfies CA; the only difference is that in the EPRB case the path of the beable will have to be retrocausal, i.e. for part of its journey this beable will be travelling backwards in time. 
 
 There are  a few possible options for responding to this argument - but most interestingly from the present perspective, we may simply choose to accept that it is likely both processes arise from the same sort of underlying mechanism, and then conclude that \emph{neither} case satisfies CA!  The main argument put forward in ref \cite{Evans_2013} to support the CA approach is that continuous spacetime paths are the \emph{`widely accepted `intuitive' picture of the explanation of quantum correlations in the SEPRB cases.'} But if we have learned anything in a century of trying to make sense of quantum mechanics it is that `intuitive' pictures aren't always a good guide to reality. The `local paths' explanation is intuitive largely because it allows us to suppose that the behaviour of quantum objects is similar to the behaviour of objects we're familiar with in the classical world, but there is no good reason to think that the quantum world looks similar to the classical world. So perhaps we should simply accept that even in the SEPRB experiment there is not really any photon travelling along a continuous spacetime path: the second measurement depends directly on the first measurement, with no need for any photon to mediate the correlations between them.  This may seem like an extreme position, but it is in fact a very natural one to adopt if we have already accepted that violations of CA are possible: we will discuss this view further in section \ref{myway}.
 
 The authors point out that the causal relationship in SEPRB can be used for signalling and control of macroscopic processes, which they seem to regard as evidence against its being unmediated. But if we accept the possibility of genuine causal influences at a distance, there seems no reason why these influences should not be used for signalling and control. Indeed, the fact that the apparently unmediated influences involved in the EPRB experiment  \emph{can't} be used for signalling has long been regarded as a puzzling feature of quantum mechanics which is in need of explanation\cite{SpekkensWood}, so clearly for at least some people in the field there is nothing conceptually problematic about the idea of unmediated influences giving rise to signalling and control. 

\section{Retrocausality within a Wider Conceptual Schema\label{AVD}}

Although the term `retrocausality' appears quite frequently in the foundations of quantum mechanics, proponents of these approaches often refrain from commenting on the nature of the causal relations involved or the broader metaphysical picture required to accommodate retrocausality. However, as noted in ref \cite{Adlamspooky}, there are two importantly different notions of retrocausality floating around in the literature.  Some approaches seem to be postulating two distinct directions of dynamical causality which together determine intermediate events by forwards and backwards evolution respectively from separate and independent initial and final states - for example, the forwards-evolving state and the backwards-evolving state in the two-state vector interpretation\cite{Aharonov}. Other approaches postulate an `all-at-once,' picture  where the laws of nature apply atemporally to the whole of history, as for example in Wharton's all-at-once Lagrangian models\cite{Wharton_2018}; in such a picture the past and the future have a reciprocal effect on one another, so there is definitely some kind of influence from the future to the past at play, but these effects can't be separated out into separate forwards and backwards evolutions. We will not address here the question of whether these reciprocal all-at-once  influences can be properly called causal - there are certainly legitimate reasons to dispute this terminology, but nonetheless we will continue to refer to all-at-once approaches as `retrocausal' in order to maintain consistency with the literature.

In this section, we will examine some different ways in these competing notions of retrocausality might be embedded within a broader metaphysical picture, and consider the consequences for the strategy of using retrocausality to rescue locality in the CA sense. We begin by discussing the possibilities for retrocausality within a Humean metaphysics, ultimately arguing that the CA principle is redundant in this context. We then move to the 
 `realist' context in which retrocausality is regarded as a form of objective modal structure, concluding that within the realist approach, `all-at-once' retrocausality has many advantages over the dynamical approach and thus it is likely that any successful realist approach to retrocausality must ultimately be based on an all-at-once picture. Finally, we will argue that the CA principle is likewise redundant within  the all-at-once picture. Thus we conclude `rescuing locality' is not an appropriate motivation for retrocausality within any viable conceptual model of retrocausality. 

\subsection{The Humean View} 
  The `Humean' approach to modality denies that there are any necessary connections in nature, including causal relations, backwards causal relations, or any more general sort of modal relation\cite{Mulder}. Humeans therefore cannot hold that retrocausal relations are real modal relations.  However, in this article we are using the term `retrocausality' without presupposing that retrocausality involves a causal relation in the usual sense, and thus it may still be possible to have something akin to retrocausality within a Humean metaphysics. For example, Humeans of the best-systems school hold that the laws of nature are simply the axioms of the systematisation of the Humean mosaic which best combines simplicity and strength (the term `Humean mosaic' refers to the set of all local matters of particular fact within a given possible world, including things like events and properties but not modal notions like causes, laws and so on)\cite{Lewis1980-LEWASG}. So proponents of this approach might be willing to say that `retrocausality' exists if the best systematization of the actual Humean mosaic is one in which some of the axioms take as input a final state and predict what happens at earlier times. 

However, for the Humean this sort of `retrocausality' can have no more than descriptive significance: it just so happens that the actual distribution of events in the Humean mosaic is such that the mosaic can be efficiently described using retrocausal notions, but retrocausal influences are not actually responsible for anything that happens in the mosaic. And likewise, the Humean will usually be able to produce laws which look local in some sense, but that locality will be purely descriptive. Provided that all correlated events within a Humean mosaic have at least one spacetime point in the intersection of their future lightcones and/or past lightcones, we will always be able to write down a CA-satisfying axiomatisation in which we postulate, as a purely formal device,  mediating beables which lie along continuous spacetime paths (including possibly some retrocausal paths). However, it will also always be possible to come up with an  axiomatisation of equal predictive power in which events depend directly on one another without mediation. And there is no obvious reason to insist on the CA-satisfying axiomatisation here, because according to the Humean the beables lying on continuous spacetime paths between events can't have any causal or nomic power to influence events along their path, and therefore the presence of these beables can't actually be necessary to bring about the relevant correlations.  Because there are no objective modal relations in a Humean universe, it is not objectively the case that the events depend directly on one another or that the correlations between them are mediated by physical processes: it is just a matter of fact that the events in question are correlated and in fact nothing whatseover brings those correlations about. 

 Thus to adopt a Humean strategy, proponents of the CA approach would have to argue that the axiomatisation with mediating beables is simpler than the approach without mediating beables - indeed, `robustly' simpler in Lewis's sense\cite{10.2307/2254396} - in order to claim that the CA-satisfying axiomatisation is in fact the `best system.' It doesn't seem obvious that such an approach is indeed always simpler, but perhaps an argument could be made based on the claim that continuity is in some sense simpler than discontinuity. Nonetheless, it's clear this is a very different argument to the one the proponents of continuous spacetime paths typically make. For the Humean the mediating beables are purely `descriptive fluff' which make the theory look simpler; they are not necessary to make the theory work, nor to guarantee consistency with relativity. Indeed locality itself has no deep significance within the Humean approach, as it is nothing more than one possible form of simplicity. So those who wish to use retrocausality to rescue locality do have the option of understanding retrocausality in Humean terms,  but this is a deflationary approach which significantly reduces the import and interest of the retrocausal models.

\subsection{The realist approach}

 The alternative view, which I will refer to as `realist,' maintains that retrocausality is a  part of the objective modal structure of reality. That is, retrocausal influences are real modal relations - perhaps causal relations, or perhaps more general modal relations such as metaphysical necessitation, ontological priority, or ontological dependence. We will now consider how dynamical and all-at-once retrocausality could be embedded in a broader metaphysical picture of modal structure.
 
 One common way of thinking about causation and other types of modal structure is in terms of \emph{`dynamic production'}\cite{chen2021governing}, which refers to a `\emph{metaphysical picture of the past generating the future}' \cite{Maudlin2002-MAUQNA}. In this picture, we suppose that modal structure plays an active role in producing the future from the present. This view of modal structure is naturally coupled with the A-theory approach to time, which tells us that past and future have different status and the present is a moving interface between the future and the past\cite{10.2307/2248314}. Dynamic production   seems to be presupposed by the approach to physics sometimes known as the `Newtonian schema,'\cite{Wharton} i.e. the idea that the universe is something like a computer which takes an initial state and evolves it forwards in time: computers literally `produce' solutions in a temporal process wherein an input state is subjected to a sequence of state transformations, and the dynamic production picture assumes that the universe works in roughly the same way. 
 
Another way of thinking about causation and other types of modal structure involves a metaphysical picture in which the objective modal structure determines what happens in the block but there is no process in which the modal structure actively generates anything:  the course of history selected by the modal structure is simply instantiated eternally and timelessly \cite{chen2021governing, adlam2021laws}. We will refer to this approach as `\emph{block instantiation}.' The block instantiation model for modality is naturally coupled with  the B-theory approach to time, which tells us that present, past and future have the same status and so we have a `block universe' which exists eternally and timelessly. However, our analysis here is concerned primarily with the role played by objective modal structure in shaping the course of history, not with the nature of temporal becoming, and therefore it is in principle possible to imagine that one might choose positions on the metaphysics of time and the metaphysics of modal structure that are not aligned in this way. For example, one could combine a block instantiation model for modality with a moving spotlight approach to time,   where the block selected by the objective modal structure exists timelessly but the present moment moves through the block\cite{Cameron2015-CAMTMS-3}. Or one could even combine a block instantiation model for modality with a growing block view, where the selection of a course of history by the laws is eternal and timeless but the actual instantiation of that course of history takes place via a growth process in accordance with the solution selected by the laws. In this article we will not be concerned with these sorts of distinctions,  as our aim is to understand the way in which retrocausality may be embedded in an objective modal structure, and it is the process of \emph{selecting} the contents of reality rather than the metaphysics of temporal becoming which is relevant to this discussion. 
 
Note that it isn't our intention to suggest that a `dynamic production' picture is particularly plausible or even coherent. In fact we consider block instantiation models to be preferable for many reasons, but it's important for us to address the possibility of dynamic production because it seems likely that attempts to use retrocausality to rescue locality are to at least some degree influenced by intuitions based on a production-style picture.

 \subsubsection{Retrocausality in dynamic production models}
 
Dynamic production is viewed with suspicion by many modern physicists, largely because of a  common assumption that  A-theory approaches to time are inconsistent with relativity, since the `present moment' or the `front of the block' seems to single out a preferred reference frame\cite{sep-time}. But in fact, as shown by Earman and Pooley, it is possible to come up with a relativistic version of  growing block models which does not single out a preferred global present\cite{10.2307/42705837,Earman2008-EARRTP}. We need only say that the `temporal becoming' process is not a total order, but rather a partial order - if two spacetime points stand in a timelike or lightlike relation to one another there is a fact of the matter about the order in which they come into being, but if two spacetime points stand in a spacelike relation to one another then there is no fact about the order in which they come into being. 

However, this approach encounters problems when we try to incorporate quantum mechanics, for if we accept that Bell inequality violations entail that the universe contains instantaneous nonlocal influences, and if we assume that these influences are causal or at least asymmetric, we then will require some spacelike separated spacetime points to come into being in a fixed order so that they are able to nonlocally influence one another. For example, in an interpretation of quantum mechanics which tells us that the choice of measurement at position $B$ in figure \ref{prep} has a direct, unmediated nonlocal influence on the result of the measurement at position $C$, clearly we need the spacetime point $B$ to come into being before or concurrently with the spacetime point $C$, so we end up with a `preferred class' of reference frames, i.e. those reference frames in which $B$ occurs earlier than or at the same time as $C$. Thus it would seem that proponents of generation-based metaphysics who wish to deny the existence of preferred reference frames must find a way to accommodate the empirical results of quantum mechanics without postulating any genuine nonlocal influences between spacelike separated events - and as we have seen, one possible way to do that is to invoke retrocausality.

 But what would  retrocausality look like within a dynamic production picture? First, we evidently cannot invoke all-at-once retrocausality. For all-at-once laws apply to the whole of history all at once, whereas in the dynamic production picture we need the laws to operate piecewise on reality during the process of production; so the dynamic production picture must be coupled with dynamical retrocausality. But there are still difficulties. One obvious issue is that   the notion of production implies ordering: if one event produces another then the first event must already exist before it can produce the second event. This works straightforwardly if we suppose that production occurs only in one temporal direction, since in that case we can insist that events are generated in some temporal sequence. But  problems arise as soon as we allow both forwards causality and retrocausality, as in models using retrocausality to explain the Bell correlations, because then the future events must produce the past events but also the past events must produce the future events, so it's unclear how the generation process can ever get off the ground. 
 
In order to solve this problem it seems likely we would have to accept a \emph{mutable timeline} metaphysics. It is common within both philosophy and wider popular culture to distinguish between time travel with mutable and immutable timelines: the former entails that if you go back in time and interfere with events you can change the past and may thus end up changing the present, whilst the latter entails that if you go back in time and try to interfere with events you will find yourself unable to change anything about the past and thus will also not change anything about the present\cite{VanInwagen2010-VANCTP-3}. In a similar way, one might distinguish between  `mutable timeline retrocausality,' where an event having a retrocausal influence on the past can change the present in virtue of changing the past, and `immutable timeline retrocausality,' where  an event in the present having a retrocausal influence on the past can't change anything about the present because its retrocausal influence has already been exerted. Allowing a mutable timeline offers several possibilities to  accommodate retrocausality within a dynamic production picture, because we can have one event generate another event which then in turn exerts a backwards influence which alters the original event, so each event in turn generates the other through some number of rounds of successive alterations. For example, we might suppose that the growing block grows to some time $t$, and then a retrocausal influence from time $t$ `reaches back' into the past and makes some alteration to events at time $t' < t$ which have already occurred, and the existing events inside the block between $t'$ and $t$ are updated to be consistent with the change (Van Inwagen proposes a model similar to this to allow for mutable timelines within a growing block universe model \cite{VanInwagen2010-VANCTP-3}). Another possible option would be to postulate two distinct growing blocks, one growing from the beginning of time and one growing from the end of time, with the former generating forwards causal influences and the latter generating backwards causal influences: at some point the blocks meet, whereupon the future block would begin to penetrate the past block and vice versa and each would effect changes within the other.

However, mutable timeline approaches lead to a number of problems. First off, like mutable timeline versions of time-travel, mutable timeline retrocausality can potentially lead to paradoxes. For example, we can imagine a retrocausal version of the grandfather paradox, in which Bob takes an action which has a retrocausal influence on the past so as to prevent Bob's grandfather from being born, meaning that Bob himself never exists to take this action in the first place. Now as explored by Price in ref \cite{PRICE_1994}, it transpires that the kind of retrocausality needed to explain the Bell correlations actually can't be used to create causal paradoxes of this kind - but this fact leaves us with further questions about \emph{why} the retrocausality involved in quantum mechanics should so neatly avoid causal paradoxes. It has been pointed out that the possible retrocausal influences must be carefully fine-tuned to achieve this\cite{almada2015retrocausal}. One possible route to explaining this fine-tuning is suggested by 
 ref \cite{almada2015retrocausal}, which demonstrates that the fine-tuning of the retrocausal influences in Schulman's retrocausal model is a consequence of a particular symmetry of the theory, i.e. the fact that for a particular function $W$ and angle $\theta$ that enter into the model, $\forall \theta  \ W(\theta) = W(-\theta)$. Explaining fine-tuning by appeal to symmetry considerations is certainly better than simply regarding it as brute fact, but it still has a slightly conspiratorial air: after all, even if Schulman's model is correct it is presumably only a contingent fact that the model happens to obey this particular symmetry, so what would have happened if the symmetry had not obtained? It appears to be only this contingent fact which prevents the model from giving rise to logical paradoxes, which seems troubling - we would really like the absence of logical paradoxes to rest on a more robust foundation. 
 
Probably the most natural way to guarantee the absence of  paradoxes would be to say that the set of allowed retrocausal influences is subject to some sort of constraint forbidding influences which could give rise to causal paradoxes. This is exactly the solution typically invoked in response to a similar dilemma within the philosophy of general relativity: general relativity admits solutions which include closed timelike curves, so in order to avoid the grandfather paradox and similar problems, it is necessary to impose consistency constraints on the fields along the curve.  Earman observes that these constraints probably can't be expressed wholly in local terms - `\emph{in general, the consistency constraints may have to refer to the global structure of spacetime'}\cite{Earman2323}  - and  this seems likely to be true in the case of retrocausality in QM also: the existence of causal paradoxes is not a property of individual events but rather a holistic property of some collection of events, and therefore constraints forbidding them will not in general  be expressible in local terms.  That is to say, the constraints in question will probably have to be global `all-at-once' constraints, which as we have observed makes them a poor fit for a dynamic production picture.

 A further problem with the mutable timeline approach is that that it leads to a particularly strong form of scepticism with respect to knowledge of the past. For as we have noted, it seems natural to understand dynamic production models in terms of the A theory metaphysics of time, where the front of the block  represents `the present' and the growth process is tied to our experience of the passage of time. This would mean that we presumably  experience a moment when the block first generates it, and therefore any retrocausal influences that go back and change the past will be too late for us to actually witness their effects. Thus, for example, if retrocausality of this kind is used to explain the correlations in Bell experiments, it follows that we will never actually observe the correlations predicted by quantum mechanics. Now this is not actually inconsistent with the empirical evidence, because the retrocausal influences will change our memories and records so we will \emph{think} that we witnessed the appropriate quantum correlations, even though in fact we did not. However, this requires us to accept that our records of the past very frequently record events which never actually took place, which places us in a somewhat parlous epistemic situation and raises a number of thorny questions around the way in which scientific confirmation would work under such circumstances. To our knowledge no proponents of retrocausal interpretations have so far raised or attempted to answer these questions, which  seems to indicate that most modern proponents of the approach don't intend to be committed to a mutable timeline version of retrocausality. Certainly these issues give us good reason to avoid mutable timeline retrocausality if possible and thus to prefer a block instantiation model over a dynamic production model for retrocausal approaches to quantum mechanics.

Moreoever, even if these dual-direction dynamical production pictures could be made coherent, they don't seem to have much to recommend them. The main argument for `growth'-type models in the philosophy of time is to  preserve the intuition that there is something special about the present - the growing block universe makes the notion of temporal becoming literal\cite{Deng2017-DENMSO-3}. But accommodating retrocausality in a generation model would require us to postulate something like two blocks growing in opposite directions, which  undercuts this argument - are there other conscious beings whose present is being carried backwards through time? Why is there no evidence for this other `present' or the relations between the two `presents'? Even in the case where we have a single block with some retrocausal influences reaching back into the block, it still seems hard to understand why `the present' is necessarily at the front of the block even though other parts of the block are also undergoing changes. There are also a whole host of more general objections to A theory approaches and growing block approaches, such as the fact that we  need to postulate a second timeline to measure the growth of the block, and this kind of `two time' view is unappealing for a number of reasons\cite{baron2018introduction}. We won't rehearse these well known arguments here, but it's worth nothing that most of these objections may also be regarded as objections to the standard `Newtonian schema' in which the universe is something like a computer: computers generate solutions in a temporal process since they exist \emph{within} the spacetime structure of the universe, but we have no clear evidence that the universe itself exists within some larger spacetime structure and thus it may not be coherent to imagine the universe as a whole being produced in a temporal sequence.

 \subsubsection{Retrocausality in block  instantiation models}

 Since it seems the pairing of retrocausality  with a dynamic production approach is an uncomfortable fit, let us now move to the block instantiation approach.  In this paradigm, laws of nature no longer `produce' the universe in some quasi-temporal process: instead they assign probabilities to entire courses of history in an all-at-once manner, and then some particular course of history is selected and the block universe instantiating that solution exists timelessly and eternally. To formalise this idea we will use the framework  introduced in ref   \cite{adlam2021laws} and developed further in refs \cite{adlam2021determinism,adlam2022operational}, which is designed to accommodate a variety of laws outside the time-evolution paradigm, including all-at-once laws. This framework construes lawhood in terms of \emph{constraints}, which are defined extensionally, appealing to techniques employed in modal logic: a constraint is defined as a set of Humean mosaics, i.e. the set of all mosaics in which that constraint is satisfied. We emphasize that the use of this terminology is not intended to reflect a commitment to the standard Humean ontology consisting \emph{only} of the Humean mosaic -  we use the phrase `Humean mosaic' to refer to all of the actual, non-modal content of reality, but since we are now focusing on realist approaches to modality, we are committed to the   claim that reality has objective modal structure above and beyond the Humean mosaic. In particular, we  will postulate that every world has some set of laws of nature which are an objective fact about the modal structure of that world, and we characterise these laws in terms of \emph{probability distributions over constraints}. For example,  a law which forbids superluminal signalling would be understood as assigning probability $1$ or close to $1$ to the set of all mosaics in which no superluminal signalling occurs. Within this picture, we can imagine that the laws of nature operate as follows: first, for each law a constraint is drawn according to the associated probability distribution, and then the constraints \emph{govern} by singling out a set of mosaics from which the Humean mosaic of the actual world must be drawn - i.e. the actual mosaic must be in the intersection of all the chosen constraints. That is to say, the laws of nature associated with a given world are understood to operate by narrowing down the set of physical possibilities for that world, thus dictating what properties the Humean mosaic for that world is required to have. 
   
  It is then straighforward to see that in such a picture, `all-at-once' retrocausality can enter  via a kind of cross-time mutual coordination. For example, in a Bell experiment where the two Bell particles are measured in the same basis, our all-at-once laws must assign probability $1$ to the set of mosaics in which the results on the two particles match, and probability $0$ to the set of mosaics in which the results on the two particles do not match. The two measurements thus have a reciprocal effect on one another, since the measurement outcomes are required to be the same if the settings are the same, but this probability distribution is completely symmetrical, so there is no sense in which one `causes' the other. Therefore if one measurement is in the future of the other we will have something that looks like retrocausality (in the general sense of `some sort of influence from the future to the past') although there is no mediating process in either temporal direction.

 But what about dynamical retrocausality? It may be tempting to say that dynamical retrocausality simply can't exist within a block instantiation model, but some care needs to be taken here, because the block universe picture is very common amongst modern physicists, and most of those physicists currently believe in the existence of dynamical causality in at least one direction. In fact, the standard way to make sense of `dynamical causality' within a block universe picture is to move from thinking in terms of a process of temporal evolution to thinking in terms of dynamically possible histories: e.g. `\emph{A \emph{history} of the system is then a smooth function $q : \mathbb{R} \rightarrow Q$ assigning to each time $t$ the configuration $q(t)$ of the system at that time. The dynamical equations of Newtonian mechanics distinguish dynamically possible from dynamically impossible histories.}'\cite{pittphilsci16622} This way of thinking about dynamical laws allows us to express them  in a block universe context using the constraint framework: a dynamical law is analysed as a probability distribution assigning probability $1$ to the set of Humean mosaics which contain only histories which are dynamically possible according to the law. 
 
So if we want to have both forwards and backwards dynamics, as in a dynamical retrocausality picture, presumably we would need two distinct sets of dynamical laws,  one assigning probabilities to sets of mosaics that we are to think of as defining `forward histories' and another assigning probabilities to sets of mosaics that we are to think of as defining `backward histories.' For a well-defined forward dynamics, we require that  that for every state $s$ and time $t$,  the probabilities assigned by the dynamics to mosaics featuring state $s$ at time $t$ sum to exactly $1$, and likewise for the backwards dynamics. In the dynamical picture we are supposed to be able to think of the laws of nature as defining the course of history based on a fixed initial state and final state, so we must select an initial state and then define a new distribution for the forward histories conditioned on that initial state; provided that the dynamics is well-defined this will give us a properly normalised probability distribution. Likewise we select a final state and then define a new distribution for the backward histories conditioned on that final state. Finally, we must  draw sets of mosaics according to the distributions associated with the laws and then select the actual course of history from the intersection of the chosen sets.  
 
 An obvious problem with this approach is that neither a mosaic or a set of mosaics is intrinsically directed; a mosaic is simply some collection of local matters of particular fact. So in this dual dynamics picture, the sets  of mosaics over which the forward-dynamics assigns probabilities are  `forward-directed' purely by stipulation, and likewise for the backward-dynamics. Imposing some requirement of `locality' also does not help  differentiate between forwards and backwards dynamics, because a mosaic where we have continuous `forwards' spacetime paths will also be a mosaic with continuous `backwards' spacetime paths. It's therefore quite unclear what it means to say that one set of dynamics is `forward' and the other is `backward': really we are just dealing with two distinct dynamics conditioned on different states. So conceptually, this `two directions of dynamical causality' looks quite difficult to sustain within a block model.

 A further problem is that   it's hard to maintain the requirement that the two dynamics are separate and independent without running into inconsistencies. For example,  if the forward dynamics say that, conditional on the actual initial state, event $E$ has probability $1$ of happening, while the backwards dynamics say that, conditional on the actual final state, event $E$ has probability $0$ of happening, then the intersection from which we are supposed to select the actual course of history will be empty. Of course, it's also possible in the all-at-once picture that the intersection of the selected sets can be empty, but within an all-at-once picture we are free to simply stipulate that the laws are such that this intersection is never empty. We don't have that same freedom in the dynamical picture because part of the premise is that the initial and final state are distinct and independent inputs from which the rest of history is generated; thus we can't simply place constraints on the combination of initial and final state such that the two dynamics are never in outright contradiction, since that would undermine their supposed independence. After all, if we accept that the initial and final state can be selected jointly in this `all-at-once' manner, then why not just allow the whole history to be selected `all-at-once' and avoid these sorts of difficulties altogether?

 This difficulty is obscured in some existing retrocausal approaches because it is common to apply the retrocausal description to just one experiment at a time, so the forwards dynamics need only meet the backwards dynamics once. For example in the transactional interpretation, we typically focus on a single experiment in which a variety of offer waves and confirmation waves interact and ultimately produce a single transaction\cite{Cramer}; the ongoing effect of the transaction on future events is not explicitly modelled. Likewise, in the two-state vector interpretation, we are typically dealing with a preparation, a final measurement, and a weak measurement in the middle whose outcome is determined by the forwards-evolving state produced by the preparation and the backwards-evolving state produced by the final measurement\cite{Aharonov}; we are not told how to model the effects of the outcome of the weak measurement on the entire future course of events. Focusing in on individual experiments makes it easy for us to isolate distinct forwards and backwards dynamics conditioned respectively on the initial and final states of a single experiment, so we can come up with simple formulas for the effect of the forwards and backwards dynamics on intermediate events, such as the ABL rule used in the two-state vector interpretation to predict the outcomes of weak measurements conditioned on an initial and final state\cite{Mohrhoff_2001}. But if we are to take the two-directions of dynamical causality picture seriously, it follows that any experiment must be embedded within a more general forwards dynamics and backwards dynamics, with the initial and final states for the experiment themselves undergoing causal influences from both the past and the future. Moreover, both the forwards dynamics to the future and the backwards dynamics to the past are affected by the result of the intermediate measurements, so the forwards and backwards histories must also be adjusted in response to whatever the result of that measurement might be. Thus we are no longer working within the simple picture where the forward and backwards dynamics meet at exactly one point and determine exactly one event: we must somehow reconcile the two independent dynamics in a consistent way all across spacetime.

 This difficulty ties into an influential criticism made by Maudlin in his commentary on the  transactional interpretation\cite{Maudlin2002-MAUQNA}. In this piece, Maudlin imagines an experiment in which a radioactive source emits a particle  at random either to the right or the left. If no particle is detected on the right after a certain time, a detector is swung around to the left in order to detect the particle on the left instead. The transactional interpretation has some difficulty here, because the second detector can send a confirmation wave back only if it is in place on the left, but this occurs only if the particle has not been detected on the right, so the existence of the confirmation wave entails that the particle will certainly be detected on the left even though the \emph{amplitude} of the confirmation wave implies only a probability of half for this event. The reason this problem arises is that the scope of the transactional interpretation is too small: it works well if we consider only a single experiment with fixed preparation and measurement devices, but when we embed that experiment in a larger causal structure where the past and future may themselves be influenced by the outcome of the experiment, it becomes very difficult to maintain the picture of two distinct and independent dynamical processes. Maudlin's argument is thus a specific version of the difficulty we have encountered in finding a way to reconcile two independent dynamical processes within a block instantiation model. The crux of the problem, according to Maudlin, is that \emph{`if the course of present events depend on the future and the shape of the future is in part determined by the present then there must be some structure which guarantees the existence of a coherent mutual adjustment of all the free variables.'}  According to Maudlin, this is also a problem not only for the transactional interpretation but also for any other retrocausal interpretation of quantum mechanics; but in fact, on examination it is clear that Maudlin's objections is a problem only for \emph{dynamical} retrocausality. For as noted in ref \cite{RevModPhys.92.021002}, in an  all-at-once model we do indeed have such a structure - the role of the all-at-once laws is precisely to mutually adjust all of the free variables to ensure consistency and adherence to all of the laws. Thus as long as we don't have distinct dynamical processes which must somehow be combined into a unified prediction,  Maudlin's problem cannot arise. (Of course all of this is fairly schematic; it would be very interesting to see a more detailed demonstration of how Maudlin's experiment would work in an all-at-once picture). 
 
Indeed, many of the other responses to Maudlin's objection also seem to be moving towards an `all-at-once' approach. Evans' response is to observe that `\emph{the tension stems from the distinctly causal notion of “generation” in Maudlin’s metaphysical picture in contrast to the “fixity”	of a unique solution in his characterisation of determinism,'}\cite{Evans2011-EVAASO-2}. That is, Maudlin's problem is dissolved if we alter the transactional interpretation to ensure `causal symmetry,' i.e. we don't have an offer wave first and \emph{then} a confirmation wave and \emph{then} another offer wave and so on, instead we achieve temporal symmetry by invoking both initial and final boundary conditions so the whole transaction is fixed eternally and atemporally. This seems to amount to a denial that offer wave and confirmation wave can be understood as separate and independent dynamical processes: the transaction must be understood within some kind of all-at-once picture. Similarly Kastner suggests removing   the offer and confirmation process from ordinary spacetime and locating it instead in a broader possibilistic space, which requires that we \emph{`abandon the idea that there is cyclic `echoing' between absorber B and the emitter.}'\cite{RK} Kastner's response thus entails that the retrocausal influence is not mediated by physical, dynamical spacetime beables, but rather by a calculation in some other space which selects which transaction is to be actualised and then applies that result `all-at-once' without any physical mediation. We don't have space here to discuss the details of these solutions, but it's clear that they're based on a similar underlying intuition that Maudlin's problem can be avoided if we move away from the picture of dynamical retrocausality and instead take up an `all-at-once' approach.

\subsubsection{Continuous Action in the `all-at-once' picture}  

Demanding CA makes sense in a  metaphysics of dynamic production, because we tend to think of the process by which the future is produced from the past as being itself local, so that the events at a spacetime point $x$ generate all the points in the future lightcone of $x$ immediately adjacent to $x$ and then those points generate the next points in their own future lightcones and so on. But we have just seen that a metaphysics of dynamic production leads to a whole host of problems when we try to combine it with retrocausality, and therefore it seems unlikely that many proponents of retrocausality would actually want to be committed to this sort of picture.  Indeed, many current proponents of retrocausality seem to favour all-at-once approaches paired with block instantiation metaphysics - see for example Wharton's `all-at-once' models\cite{Wharton_2018}, and Evans' work on retrocausality in a block universe\cite{Evans2011-EVAASO-2}. Moreover, many of these modern proponents want to use \emph{this sort} of retrocausality to rescue CA - but in a metaphysics \emph{not} based on a generation process, it's much less clear why we would expect our models to obey CA in the first place. In this section  we argue that in fact mediating beables are redundant within the all-at-once picture and thus `rescuing locality' cannot be the main motivation for retrocausality of this kind.

Heuristically, it's straightforward to see the problem for intermediate beables in an all-at-once approach. For in the all-at-once picture, probabilities are assigned to entire histories, which, if we zoom out far enough, would ultimately encompass entire Humean mosaics. Thus the existence or nonexistence of beables occupying the spacetime regions between relevant events in the mosaic is not really relevant, because if probabilities are assigned directly to the entire distribution of events across the mosaic, there is no need for information about one part of the mosaic to be carried to other parts by a physical beable propagating either forwards or backwards in time. After all, in this picture what happens at a spacetime point is not determined by the information locally available at that spacetime point but rather by the probability distribution assigned from the outside, which necessarily contains information about events taking place at other locations in the mosaic: our models can be as retrocausal and non-local as we like without any need for mediation, as we can simply choose probability distributions which give rise to nontrivial correlations between the relevant events. For example, in the Bell experiment shown in figure \ref{prep}, if we are working in an `all-at-once' model we don't need a local physical process propagating from $C$ to $B$ via either $A$ or $D$ in order for the measuring device at $B$ to `know' the choice of measurement at $C$; we simply postulate `all-at-once' laws which assign the appropriate quantum-mechanical probabilities  to mosaics exhibiting various  different combinations of measurement outcomes at $B$ and $C$.

Now, it is of course true that within an all-at-once picture we might sometimes be able to find a `local time evolution' interpretation of a given model if the probability distribution over mosaics defined by the model is factorisable in the following sense. Let us decompose reality into a set of discrete events taking place at different spacetime locations. Next, for each location we use our model to define a probability distribution over the possible events that could occur at that location, conditioned only on events  in the past lightcone of that location. Then, given some set of initial events (i.e. an initial condition) we can try to calculate the correct probability distribution over mosaics by combining all the local conditional probabilities: $P(e_1 e_2 e_3) = p(e_1) p(e_2 | e_1) p(e_3 | e_2 e_3)$ where $e_1$ is in the past lightcone of $e_2$ and $e_2$ is in the past lightcone of $e_3$, and so on until we have covered the entire mosaic. If the probability distribution thus obtained matches the full distribution over Humean mosaics induced by our model, then that model can be understood in terms of a `local time evolution' picture where we can think of the laws locally generating each step based on information in the local (past-determined) state at that spacetime point. If it isn't possible to create a distribution in this way which matches the full distribution induced by the model, then the model can't be interpreted in terms of local time evolution.

 Thus in an all-at-once picture, to say that the laws involve `nonlocality' is simply to say that the local time evolution decomposition isn't always possible: the probability distribution over mosaics induced by the laws can't be completely decomposed into local probabilities conditioned on past lightcones. Under these circumstances, we can no longer imagine that the laws locally generate each step based on information in the local past-determined state at that spacetime point. Moreover, because assigning probabilities to entire mosaics does not give rise to intrinsically directed evolutions or causal relations, for any pair of events whose correlations can't be understood in terms of a local time-evolution decomposition, we must simply understand those events as standing in some sort of  reciprocal dependence relation. So as long as the events are at different times there will necessarily be some sort of `retrocausality' involved (in the general sense of `some sort of influence from future to past.') Thus in the all-at-once picture,  to say that  laws are `retrocausal' is simply to say that we have a nonlocal model in which some of the correlations  which can't be decomposed into local past-determined probabilities involve events at different times. And in fact this will always be the case, because the only way that two events can be at the same time in all reference frames is for them to be at one and the same spacetime point, in which case there can't be nonlocality involved. So in the all-at-once picture, to say that the mosaic is `retrocausal' just is to say that it is `nonlocal.'

Of course, one could also look for a `backwards local time evolution' model involving a similar decomposition with conditional probabilities defined over events in the \emph{future}  lightcones of the relevant points. This would allow us to give an interpretation which is local and retrocausal in the sense that the direction of causality is everywhere reversed. But the approaches we have discussed which aim to use retrocausality to rescue locality require both forwards and backwards causality, so this can't be what proponents of such models have in mind. So can we have a `forwards \emph{and} backwards local time evolution' interpretation? This would correspond to a decomposition with the probabilities for a given event conditioned on events which can be reached from that event by timelike or lightlike pathways going forward and/or backwards in time. It is  clear that Bell correlations will always have such an interpretation, since such correlations require a preparation event in the past lightcone of both measurements and hence there is necessarily an appropriate path between them via that preparation event. Moreover, any correlations that we can actually measure must have such an interpretation, since in order for us to calculate correlations between events we must be able to compare the results at some point in the future lightcones of both events, and thus there must be an appropriate path between them via that future point. So the existence of this decomposition is trivial for all correlations that we ever have measured or ever will measure, and  therefore we have no grounds for attaching a special significance to the possibility of coming up with such a decomposition: finding a `forwards local time evolution' decomposition does seem to tell us something interesting about the nature of the underlying processes, but finding a `forwards and backwards local time evolution' decomposition tells us nothing at all.

That is to say, we can have mediating beables traveling along continuous paths within an all-at-once picture if we so wish, but they are entirely redundant, since the existence of a `forwards and backwards time evolution' decomposition is necessarily trivial. Kastner makes a similar point in her article arguing that there is no real retrocausation in retrocausal approaches to quantum mechanics\cite{doi:10.1063/1.4982766}. Kastner takes the view that retrocausal approaches which include both an initial and final condition necessarily lead to a block universe metaphysics, since `\emph{their explicit time symmetry forces a static ontology}.' Moreoever, by `retrocausation' she means approaches in which the retrocausal influences are understood in terms of `\emph{dynamical propagation ... occurring in spacetime'} i.e. what we have referred to as dynamical retrocausality. Now, we have seen in this article that retrocausal  approaches don't necessarily have to be combined with a block universe metaphysics (since we have the option of a dynamic production using a mutable timeline approach). However, we have argued that there are very good reasons to prefer the block instantiation model combined with an all-at-once approach, and thus Kastner's comments still have bite. As she notes, if our background metaphysics is a block universe, all-at-once picture, then `\emph{there is no influence propagating `anywhen' in spacetime; it is just a story tacked on to a set of events that already exist ... it is the static block world that is doing the work of `saving locality,' not any dynamical process.}'

Indeed, the foregoing discussion makes it clear that in an all-at-once picture there is a sense in which locality is very \emph{unnatural}: if we are dealing with laws which assign probabilities to entire mosaics, then generically it \emph{won't} be possible to come up with a forwards local time evolution decomposition of these laws, so laws which are `local' are actually very special cases. So if we believe in the `all-at-once' picture the existence of nonlocality is not in the least surprising - the real surprise is the fact that it is so limited!

\section{Retrocausality from Nonlocality \label{myway}} 

We have so far made the argument that while it is technically possible to use retrocausality to save locality, there is no very good reason to do so. In this final section, we will make the complementary argument: in fact accepting the existence of genuine, unmediated nonlocality in and of itself leads us to accept retrocausality. More specifically, if we attempt to give a realist, intersubjective account of the  Bell correlations in terms of some mechanism other than beables lying along a continuous spacetime pathway, and we are not willing to accept the existence of a preferred reference frame, then we have good reason to accept the existence of retrocausality in the unmediated all-at-once sense.

First, observe that as argued in ref \cite{Adlamspooky} it follows straightforwardly that if we accept the existence of an unmediated influence between spatially separated measurements in a spacelike Bell experiment, then we must also accept the existence of an unmediated influence between temporally separated measurements in a Bell experiment, since the spatially separated measurements become temporally separated measurements under a change in reference frame. This conclusion could be avoided by postulating that the reference frame in which the measurements are spatially separated is a preferred reference frame, but we have committed to avoiding preferred reference frames so this option is not available to us. Thus, the existence of spatial nonlocality together with a commitment to avoiding preferred reference frames inevitably leads us to the possibility of nonlocality not merely in \emph{space} but in \emph{time}. 

Moreover, this nonlocality in time will also have a retrocausal nature in some frames of reference. For example, suppose we postulate that there is an influence from $B$ to $C$ in figure \ref{prep}. But there are many reference frames in which $C$ is in the past of $B$, so in those reference frames this influence will be retrocausal. Clearly the same is true if we say the influence is from $C$ to $B$ instead, or indeed if we say that it goes both ways: there is no relativistically covariant way to guarantee that the influence will always go forward in time. 

Now,  this demonstrates the existence of temporal nonlocality and retrocausality only between spacelike separated events, and one might well feel that since spacelike separated events have no well-defined temporal order this doesn't count as retrocausality in any strong sense. We can of course rearrange our experiment such that $C$ is in the future lightcone of $B$, but in that case it is no longer strictly necessary for us to postulate any kind of nonlocality, since we can always suppose that there is some physical signal travelling from $B$ to $C$. However, this way out entails accepting that the underlying mechanism by which the correlations are brought about changes when we move $B$ into the future lightcone, even though the correlations are exactly the same. This contravenes the common sense notion sometimes known as `ontic equivalence,' i.e. the idea that if two processes produce the same operational statistics, they should where possible be attributed to the same kind of underlying mechanism\cite{Spekkens}. There are cases within quantum mechanics where it appears that we have to violate ontic equivalence (this is essentially the content of the contextuality theorems\cite{Spekkens}) but this does not appear to be one of those cases - the asymmetry can easily be removed by supposing that the same nonlocal mechanism operates in both cases, and therefore it seems we have good reason to make that supposition. Indeed, since the particle at point $B$ has no way to know whether or not $C$ is in its future lightcone at the time of the measurement unless the particle at position $C$ sends some sort of retrocausal signal back to position $B$, it follows that if we want to avoid timelike retrocausality we would have to suppose that when $B$ is measured it \emph{both} sends a nonlocal influence directly to $C$ and \emph{also} broadcasts a physical signal into the whole of its future lightcone just in case $C$ is to be found there; and then since we know that the nonlocal influence is sufficient to produce the observed correlations, this physical signal seems superfluous, unless we are supposed to imagine that simply being in the future lightcone of $B$ somehow blocks the particle at $C$ from being affected by this nonlocal influence. Moreoever, once we have accepted the existence of nonlocal influences across spacelike separations, we have already admitted such influences into our ontology and therefore there seems much less reason to object to the possible existence of such influences across timelike separations.

So far this argument gets us only to the existence of temporally non-local influences across timelike separations: we have not shown that these influences must be retrocausal. However, returning to the spacelike separated case, observe that  if the particles employed in the experiment are maximally entangled, there's no operational difference between them and so we can't appeal to the statistics to determine which measurement `causes' the result of the other. Nor can we appeal to time-ordering, since there is no relativistically covariant way to determine time-ordering in the case of spacelike-separated events. We can't even choose the direction of causality at random - any choice of probability distribution would look biased in some reference frames. So it seems that if we are determined to avoid preferred reference frames we can't hold that one measurement causes the result of the other and not vice versa. Instead we must maintain that the two particles undergo a kind of  non-local joint coordination - their correlations are fixed `all-at-once' rather than in some particular sequence, so the effect of the particles on one another is perfectly reciprocal. The same conclusion is reached in ref \cite{almada2015retrocausal}, which argues that choosing one of the measurements to be the cause of the other breaks the operational symmetry of the scenario, and argues that `\emph{If this broken symmetry is restored,
	where influence is always allowed to go both ways (no matter when the measurements
	occur), then these models become retrocausal as well as superluminal.}'

Finally, since we have already argued that if possible we should suppose that the same mechanism is in operation when the measurements are performed at a timelike separation, it follows that when $C$ is in the future lightcone of $B$ the correlations still arise from non-local joint coordination and thus the particles have a reciprocal effect on one another - that is, $B$ has an influence on $C$ but also $C$ has an influence on $B$, so we do indeed have retrocausality. This form of retrocausality, however, is not mediated via any beables lying along continuous spacetime pathways: it is a natural consequence of the kind of `all-at-once' picture that we discussed in section \ref{AVD}. (As we noted earlier in the paper, one might well feel that this kind of all-at-once, reciprocal influence is not really `causal' in nature - we are not committed to the claim that it is `causal' in any strong sense, but since it   involves some kind of backwards in time influence, it qualifies as `retrocausal' in the very general sense in which we have been using that term).

We summarise the structure of this argument now: 

\begin{enumerate} 

\item Suppose that there are no preferred reference frames in the universe (either observable or unobservable)  - i.e.  there is relativistic covariance at all levels of reality
	
	\item Suppose that when an EPR experiment as in fig  \ref{prep} is performed with $B$ and $C$ at a spacelike separation, the correlations are explained by a non-local influence between $B$ and $C$

	\item There is no relativistically covariant way to decide whether the nonlocal influence in an EPR experiment goes from $B$ to $C$ or vice versa

\begin{enumerate}

	\item 1), 2) and 3) together entail that in an EPR experiment there is a  nonlocal influence between the measurements at $B$ and $C$ which is \emph{reciprocal}  - i.e. the influence is from $B$ to $C$ and also from $C$ to $B$
	
		\end{enumerate} 
	
	\item Suppose the ontic equivalence principle is valid: where possible,  two processes which produce the same operational statisticsshould  be attributed to the same kind of underlying mechanism
	
	\item The operational statistics for an EPR experiment as in fig \ref{prep}  are the same regardless of whether $B$ and $C$ are at a spacelike separation or $C$ is in the future lightcone of $B$
	
	\begin{enumerate}

	\item 4) and 5) together entail that we should assume the underlying mechanism responsible for the EPR correlations is the same regardless of whether $B$ and $C$ are at a spacelike separation or $C$ is in the future lightcone of $B$ 
	
	\end{enumerate} 
	
	\item 3a) and 5a) together entail that  when an EPR experiment  as in fig \ref{prep} is performed with $C$ in the future lightcone of $B$, the measurements  have a reciprocal influence on one another. 
	
	\item 6) entails that when $C$ is in the future lightcone of $B$, it has a `retrocausal'  influence on the measurement at $B$; so timelike retrocausality exists

	\end{enumerate} 

This argument gives us reason to believe in the existence of timelike retrocausality in the case of entangled particles; in and of itself it doesn't entail that retrocausality is in action for other sorts of systems. However, someone who accepts the conclusion of this argument might well think seriously about allowing retrocausality in other contexts as well. First of all, one major motivation for avoiding retrocausal models in science is the conviction that retrocausality is the kind of thing that simply can't happen, either because it's logically impossible (as it leads to logical paradoxes, or it is ruled out by the definition of causality) or because it's nomically impossible (because bringing about retrocausality is not a nomic power which exists within our world). But if we accept that there is timelike retrocausality in the case of entangled particles, we must acknowledge that  retrocausality (in the general sense in which we have used the term here) is both logically and nomically possible, and thus one major obstacle to postulating retrocausal models has been removed. 

Second, we've already argued that the `all-at-once' approach to retrocausality is more coherent than dynamical retrocausality approaches, and moreover given the reciprocal nature of the dependence we've postulated for entangled particles, it seems the behaviour of the entangled particles is most naturally understood in an `all-at-once' picture. But it's difficult to imagine how one could possibly combine an `all-at-once' model for entangled particles with a standard `forwards time evolution' picture for everything else, because entangled particles are not separable from the rest of reality: the behaviour of entangled particles is determined by classical objects (since we prepare entangled particles with classical devices) and in turn the entangled particles affect classical objects (since we measure entangled particles with classical devices). So we can't do the `all-at-once' calculation for entangled particles without first knowing about the behaviour of various classical objects  throughout history, but we can't get to the end of the evolution of the classical objects without knowing where the `all-at-once' calculation leaves the entangled particles. Thus the natural solution is to assume that the `all-at-once' approach applies to the entire contents of reality, meaning that retrocausality applies to everything, even if it's less noticeable for classical objects. The specific way in which retrocausality is manifested for systems other than entangled particles would depend on the details of the particular retrocausal model under examination, and since most existing retrocausal models take classical objects as primitives they don't yet seek to address this question, but nonetheless this way of thinking does lead naturally to the conclusion that retrocausality (in the generalised sense used throughout this article) is probably a generic feature of reality. 

Third, we can make a version of the argument of ref \cite{Evans_2013}  discussed in section \ref{symmetry}, flipping the conclusion as suggested in that section to maintain that in fact even the straightforward single-photon experiment discussed in that paper should be thought of as involving nonlocality and retrocausality. Moreover, it's in fact generically true that EPR-style experiments involving entangled particles can be translated into prepare-transform-measure scenarios which give rise to the same operational statistics: this is a consequence of the Choi-Jamiolkowski isomorphism and the operational interpretation of it set out in refs \cite{Leifer22} and \cite{e22091063}. So if we accept the ontic equivalence principle, and we accept that experiments involving entangled particles involve nonlocality and retrocausality, we have good reason to accept that a large variety of prepare-transform-measure scenarios in quantum mechanics also involve nonlocality and retrocausality.

\section{Conclusion} 

In this article we have argued that using retrocausality to restore locality is the wrong motivation; including mediating beables in our retrocausal model does not seem to bring any concrete benefits and in any case such continuous paths are redundant both within the Humean approach to modality and within the all-at-once approach to retrocausality which seems most plausible and coherent for realists about modality.

Given the importance of modal structure for this analysis, it's worth remarking that the community's continued insistence on mediating beables may arise in part because proponents of retrocausality have an insufficiently robust approach to modality: those who are uncertain about the existence of modal structure as distinct from physical reality may feel that dependence relations need to be made physical via continuous spacetime paths or else they can't be `real.' But if there is no such thing as objective modal structure then the entities moving along these continuous spacetime paths can't have any modal or nomic power to make anything happen in any case, so the continuous spacetime paths don't actually help the situation for sceptics about modality. If you don't believe in objective modal structure then there can be no need for local physical processes to mediate modal relations; and if you do believe in objective modal structure then there seems to be no obvious reason to demand that it must always be mediated by local physical processes.

Let us finish by making some concrete suggestions for the retrocausal research programme in light of the conclusions reached here. One important point is that proponents of retrocausality might actually find that their positions look stronger if they stop making arguments based on locality. For the  evident conflict between the `continuous path' picture and the `all-at-once' metaphysics needed for a reasonable account of retrocausality may contribute to an impression of philosophical incoherence; abandoning continuous paths therefore has the potential to make retrocausality more appealing to a wider audience.

Another important consequence of our argument is that the retrocausal research programme could benefit from a shift in focus.  Up until now retrocausal models have largely focused on modelling the retrocausal process - for example, in terms of offer waves and confirmation waves in the transactional interpretation, or the interaction of a forwards and backwards evolving state in the two-state vector interpretation. But in view of the conclusions of this article, it may be that focusing entirely on the nature of the retrocausal process is something of a blind alley - if the `process' is actually a pure modal relation, it is not really located in spacetime and therefore puzzling over how it could be located in spacetime isn't the right question to ask. Note that we don't mean to suggest that models exhibiting continuous action should be actively avoided; in any explicit retrocausal model it will be necessary to formalise the modal relationships between different events in some way or other, and it may well turn out that the easiest way to do so is to write down models based on continuous paths through spacetime, with the proviso that those paths are not to be taken entirely literally. But we do contend that demonstrating consistency with continuous action should not be the main priority for the retrocausal research programme.

So what \emph{should} the priorities be? Well, as noted in section \ref{ontology}, most existing retrocausal models take macroscopic objects as primitives, but clearly at some point we'd like to understand how these macroscopic objects arise from fundamental ontology. Tackling this question may potentially lead to interesting new directions for the retrocausal research programme. One retrocausal approach which does attempt to address this sort of question is Kent's `solution to the  Lorentzian quantum reality problem'\cite{Kent,2015KentL,2017}. Kent doesn't use the term `retrocausal,'  but his models are certainly retrocausal in the general sense in which we've used the term in this article, since they tell us that the content of reality depends on the result of a  `measurement' at the end of time.  This approach offers an ingenious solution to the problem of establishing an ontology: effectively, events occur if and only if there is a record of them in the initial and/or final state of the universe. In particular, since measuring instruments are macroscopic objects, the result recorded by a measuring device quickly spreads out into the environment by means of decoherence and thus it can't realistically be erased, so there is always a record of a measurement result in the final state of the universe and therefore measuring instruments and measurement results feature in the ontology in a straightforward way. We don't have space to examine this approach in detail in this article, but whether or not Kent's proposed solution is the right one, we contend that he is addressing the right problem.

In a similar vein, in our view the strongest argument  in favour of local theories, as described in section \ref{reasons}, is the idea that postulating locality at the fundamental level helps to explain the appearance of locality in the classical world and/or the fact that the principle of locality has been scientifically successful. But we have seen that postulating locality in the CA sense \emph{together  with retrocausality} undercuts that explanation, since retrocausal models are designed to produce the appearance of nonlocality and thus in such models the locality at a fundamental level no longer explains the general appearance of locality in the classical world: that work is done by the constraints the retrocausal model places on the allowed retrocausal effects. Moreover, a very similar criticism will apply to any other approach which attempts to explain the appearance of macroscopic locality by rescuing locality at a fundamental level, such as superdeterminism - any such approach will have to come up with some method of producing the appearance of nonlocality in  Bell inequality violations, and will therefore still have to explain why this method can't be used to produce much more widespread nonlocality.  This suggests that it is a crucial problem both for the retrocausal research programme and for other interpretative approaches to explain why the classical world largely looks local despite the existence of mechanisms to produce correlations which are nonlocal in the Bell sense. Many current retrocausal models have no real explanation for this feature - they simply adopt some mathematical machinery from quantum mechanics which automatically has the consequence that nonlocality is limited in specific ways, but they make no attempt to explain why this particular machinery is suitable. Obviously, as ref \cite{hance2021wavefunction} points out, standard quantum mechanics doesn't explain this either and therefore other interpretative approaches are not doing any worse if they fail to explain it, but that seems like an inadequate response - the point of working on interpretations of quantum mechanics is precisely to explain things that standard QM fails to explain.  Our view is that explaining the appearance of locality in the classical world and QFT is a more urgent and interesting question than attempting to rescue locality with retrocausality. 

\section{Acknowledgements} 

Thanks to participants in the `Quantizing Time' workshop, and particularly Ken Wharton, for discussions which inspired some parts of this article. 

\vspace{2mm}

This publication was made possible through the support of the ID 61466 grant from the John Templeton Foundation, as part of the “The Quantum Information Structure of Spacetime (QISS)” Project (qiss.fr). The opinions expressed in this publication are those of the author(s) and do not necessarily reflect the views of the John Templeton Foundation.

  \bibliographystyle{unsrt}
 \bibliography{newlibrary12}{}

\end{document}